\documentclass[10pt,a4paper,fleqn]{article}
\usepackage[utf8]{inputenc}
\usepackage{amsmath}
\usepackage{amssymb}
\usepackage{graphicx}
\usepackage[colorlinks]{hyperref}
\usepackage[left=2cm,right=2cm,top=2cm,bottom=2cm]{geometry}
\usepackage{authblk}

\usepackage{cite}
\usepackage{xcolor}
\begin{document}

\title{Distribution of interspike intervals of a neuron with inhibitory autapse stimulated with a renewal process}
\author[1]{O. Shchur}
\author[2]{A. Vidybida}
\affil[1,2]{Department of Synergetics, Bogolyubov Institute for Theoretical Physics of the National Academy of Sciences 
of Ukraine, Kyiv, Ukraine
}
\affil[1]{Laboratory of Biological Computation, Institute of Experimental Medicine,
Budapest, Hungary
}
\affil[1]{olha.shchur@bitp.kiev.ua, shchur.olha@koki.hu}
\affil[2]{vidybida@bitp.kiev.ua, vidybida.kiev.ua}

\maketitle
\begin{abstract}
In this paper, we study analytically the impact of an inhibitory autapse on neuronal activity. In order to do this,
we
formulate conditions on a set of non-adaptive spiking neuron models with delayed feedback inhibition, 
instead of considering a particular neuronal model.
The neuron is stimulated with a stochastic point renewal process of excitatory impulses. 
Probability density function (PDF) $p(t)$ of output interspike intervals (ISIs) of such a neuron is found
exactly without any approximations made. It is expressed in terms of ISIs PDF 
for the input renewal stream and ISIs PDF for that same neuron without any feedback.
Obtained results are applied to a subset of neuronal models with threshold 2 when the time intervals between input impulses are distributed according to the Erlang-2 distribution. 
In that case we have found explicitly the model-independent 
initial part of ISIs PDF $p(t)$ defined at some initial interval $[0;T_2]$ of ISI values. 
\end{abstract}


\section{Introduction}

The brain consists of neurons that are wired together by axons with synapses and communicate
with each other by electrical impulses, or spikes.
There are two classes of neurons:  excitatory and inhibitory.
To execute cognitive tasks, in the cortex, the balance between excitation and inhibition
must be maintained \cite{Isaacson2011, Atallah2009}. Recently, the importance of cortical disinhibition, that is the temporary ceasing of inhibitory neurons activity,  was recognized,  for  learning and memory \cite{Letzkus2015}, sensorimotor integration \cite{Lee2013}, locomotion \cite{Fu2014}, social behaviour \cite{Marlin2015}, and attention
\cite{Sridharan2014}.
Disinhibition can be achieved in different ways, for instance, by neuromodulation \cite{Fu2014, Marlin2015}, long-range inhibitory input \cite{Zhang2014}, and 
also at the level of local circuits \cite{Burton2017}. In the latter case,
for some inhibitory neurons, their disinhibition  is accomplished
due to feedforward inhibition obtained from other neurons \cite{Pfeffer2013}. On the contrary, for
parvalbumin-expressing (PV) inhibitory neurons, the main source of their inhibition is autaptic transmission \cite{Deleuze2019}. 
The latter means that PV neurons send synaptic connections not only to other cells, but also to
themselves. Such inhibitory synapses are called autapses.

Most neurons in the brain are stimulated with sequences of spikes (spike trains) that appear
random \cite{Liley1956, Drongelen1978, Shadlen1998, Baddeley1997, Maimon2009, Shinomoto2009, Mochizuki2016}. Mathematically such sequences are usually described as
some stochastic point process \cite{Johnson1996}. Often the most simple case of a point process, a Poisson process, is used to describe a neuronal activity. For a Poisson process,
its intensity, or infinitesimal rate at which events are expected to occur around a particular time,
is constant \cite[Sect. 2.2]{Johnson1996}. It does not depend on the prior history of the point process. 
%It leads for the distribution of interspike intervals (ISIs,  times between two successive spikes in a train) to have exponential distribution. 
In some cases, the description of neuronal activity with a Poisson process is
experimentally approved \cite{Liley1956, Drongelen1978, Shadlen1998}, but in many others both experimental
data \cite{Baddeley1997, Maimon2009, Shinomoto2009, Mochizuki2016} and theoretical considerations \cite{Softky1992} exclude a possibility
for a neuronal activity to have a Poisson statistics.

Although theoretical studies of neurons with autapses are abundant,
most of them employ a numerical approach, e.g. \cite{Herrmann2004,Hashemi2012,Wang2014,Yilmaz2016,Guo2016,Guo2016a,Zhao2016}.
Analytical, mathematically rigorous results regarding neuronal activity are sparse and were
obtained for some neuronal models if stimulation is deterministic, for instance, constant input current \cite{Li2019}. Only considered type of stochastic stimulation in such analytical studies is a stochastic point Poisson process \cite{Vidybida2009,Vidybida2013}.
This is not surprising. For a Poisson stream, the point process intensity
is always the same. It does not depend on the previous history of the process realization.
This property considerably simplifies reasoning and calculations.
Another situation is with a general renewal stochastic process.
In this case, unlike for a Poisson process, the process intensity does depend on the prior history of the point process, namely, on the time at which the last event occured \cite[Sect. 2.1]{Johnson1996}. 
It is well known that in any problem regarding stochastic point processes,
replacing Poisson process with a general renewal one
 presents 
fair analytic difficulties \cite[p. 277]{Johnson1996}, \cite[p. 5-6]{Cox1962}.

In the current paper,
we obtain analytical results for a range of neuronal models when stimulating process is non-Poisson, but renewal stochastic process.
  The emphasis of this work is made on the role of an inhibitory autapse, or delayed inhibitory feedback. It means that we take as given the probability density function (PDF) $p^0(t)$ of ISIs of that same neuron without feedback when it is stimulated
with a renewal stream of excitatory impulses distributed randomly
according to its ISIs PDF $p^{in}(t)$, see Fig. \ref{neurons}, upper panel. Here $p^{in}(t)dt$ and $p^0(t)dt$ give the probability to obtain, respectively, the input and output ISI duration in the interval $ [ t; t+dt [ $. 
%In present paper, $t$ denotes the time passed after the start of ISI, or spike generation.
Having $p^{in}(t)$ and $p^0(t)$, we calculate the PDF $p(t)$ of output ISIs for the neuron with delayed inhibitory feedback, see Sect. \ref{gen_exp}. In this paper, we consider a {\em Cl}$^{-}$-type
fast shunting inhibition, mediated via GABA$_A$ receptors, see Sect. \ref{class}. 

We calculate exactly the distribution of interspike intervals for a neuron with inhibitory autapse  
not for one particular model, but rather for a set of non-adaptive spiking neuronal models specified in Sect. \ref{class}, which includes a leaky integrate-and-fire model. The obtained general expression for the output ISIs PDF $p(t)$
is checked   by numerical simulation of a stochastic process, see Fig. \ref{examples}.
Since the description of ISIs with a Gamma distribution is widely used in theoretical and experimental studies \cite{Lansky2016},
obtained in this paper results are applied to the case of a neuron with threshold 2 when the time intervals between input impulses are distributed according to the Erlang-2 distribution, see Sect. \ref{sect_erlang}. We have shown that then the ISIs PDF $p(t)$ has model-independent 
part, when the probability of some ISI durations does not depend on the neuronal model, and it is found explicitly, see Sec. \ref{sect_erlang}. 

Throughout the calculations, we expect that neuronal activity has settled down onto a stationary
regime. This is not necessarily the case that for any neuronal model and initial conditions 
stochastic activity approaches a stationary regime in the course of time. In the Appendix \ref{app},
we specify a condition on the model, see (\ref{f(s)condition}), which guarantees the approach,
and find exact characteristics of that stationary regime, which are used in the main part of the paper.

\section{Methods}
\label{class}
\begin{figure}
\centering
\includegraphics[width=0.9\textwidth]{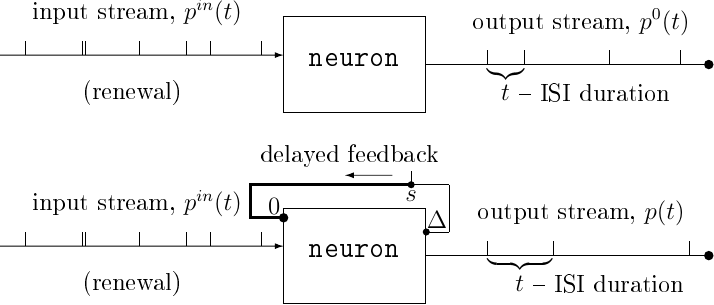}
\caption{\label{neurons} Upper panel: Neuron without feedback. Lower panel:
Neuron with delayed feedback. The neuronal model, labeled as {\large\tt neuron} in the both panels of the figure,  satisfies the description given in Sect. \ref{class}. In both cases, the neuron is
stimulated with the same stream of excitatory impulses which is a realization of some stochastic point renewal process. Thus time intervals between input impulses are independent and randomly distributed according to the distribution
$p^{in}(t)$, where $t$ is an ISI duration. As a result of the stimulation with the renewal stream, the neuron without feedback (upper panel) produces the series of spikes, which are distributed according to the
distribution  $p^0(t)$. Under the same stimulation, that same neuron, but with delayed feedback inhibition with the  time delay $\Delta$ of the feedback (lower panel), generates spikes with the ISIs PDF $p(t)$. In case of the feedback line presence (lower panel), when the neuron generates spike, 
that spike also enters the feedback line if it is empty.}
\end{figure}

In this paper, we study a set of deterministic neuronal models with 
delayed inhibitory feedback under stochastic stimulation. The neuron transforms the input point renewal stream of excitatory impulses into the output stream, see Fig. \ref{neurons}. The distribution of the intervals between input impulses is  $p^{in}(t)$, where $t$ is a time between two consecutive impulses in the input stream. 
The neuron can be triggered only at the moment of
receiving an input impulse. Further, just after triggering, the neuron fires a spike and immediately appears in its resting state, and remains there until an input impulse is received.

Since we consider neuronal models with inhibitory autapse, 
every time the neuron generates a spike, the spike also enters the feedback line, if there is no  impulse in the feedback line yet, see lower panel of Fig. \ref{neurons}. 
This additionally means that
the feedback line can bear no more than one impulse at the same time. 
The impulse in the feedback line needs $\Delta$ time units to enter the neuron. 
Taking into account all mentioned feedback properties, at the beginning of an ISI, 
which is the moment of most recent firing,  there is always an impulse in the feedback line.  
If the line was empty just before the most recent firing then the impulse enters the line due to
that firing. Otherwise, it has entered the line due to a previous firing and still has not
passed through the line.
The time an impulse in the feedback line needs to reach the neuron is called time-to-live and denoted as $s$, see the lower panel of Fig. \ref{neurons}.
Inspired by autapses of PV neurons, 
which have GABA$_A$ receptors  \cite{Bacci2003}, we assume that the inhibitory impulse from the feedback line immediately returns the neuron to its resting state.  After that, it does not have any influence on the neuron. 
Note that the case of instantaneous inhibitory feedback (i.e. $\Delta=0$) 
corresponds to a neuron without any feedback for the neuronal models studied in this work.
More details on the biological justification of considered feedback inhibiton properties can be found in \cite{Vidybida2018}.

The main goal of the present work is to derive the distribution $p(t)$ of 
intervals between impulses generated by the neuron with inhibitory
autapse, where $t$ denotes an ISI duration, see lower panel of Fig. \ref{neurons}. It is important to emphasize that we aim to
express $p(t)$ in terms of the feedback delay $\Delta$, distribution of interspike intervals for the input stream $p^{in}(t)$ and the output stream in case of feedback absence, $p^0(t)$.

The neuronal models that satisfy mentioned in this section conditions
include a perfect integrator \cite{Burkitt2006}, a leaky integrate-and-fire neuron \cite{Burkitt2006}, and a binding neuron \cite{Vidybida2014}.
Another models which may fall into the considered set belong to the renewal (non-adaptive) neurons as defined in
\cite{Gerstner2014}.

\section{Results}
\subsection{General expression for output ISIs PDF $p(t)$}\label{gen_exp}
We study a neuron with delayed feedback inhibition specified in Sect. \ref{class} in its stationary regime. 
%Also, it is worth to mention that when a neuron fires, an ISI starts anew.
Then the ISIs PDF $p(t)$ for such a neuron can be calculated as follows:
\begin{equation}\label{pt}
p(t)=\int\limits_0^\Delta ds\: p(t|s) f(s).
\end{equation}
Here  $p(t|s)$ is  the conditional probability density
to get the output ISI of a duration $t$ if at the beginning
of the ISI there was an impulse in the feedback line with
time-to-live $s$. 

As regards $f(s)$, it is the PDF  of times-to-live $s$ 
at the beginning of an ISI, i.e. just after spiking. We consider a stationary regime here.  
In the Appendix \ref{app}, it is rigorously proven that if the condition (\ref{f(s)condition}) on the PDF for the neuron without feedback $p^0(t)$
and the time delay of the feedback $\Delta$ is met, then in the course of neuronal activity $f(s)$  converges to a unique stationary distribution. 
When the stationary distribution is attained, $f(s)$ has the following form:
\begin{equation}\label{fs}
f(s)=g(s)+a\:\delta(\Delta-s).
\end{equation}
Here $g(s)\in C([0;\Delta])$ is a solution of Eq. (\ref{geq}), which is given by the expression   (\ref{g}), see the Appendix \ref{app} below. The coefficient $a \in [0;1]$ can be found from the normalization condition on $f(s)$. Notice that the expression (\ref{fs}) is in a sense universal.
The same form of $f(s)$ has been used in \cite{Vidybida2009, Vidybida2015, Vidybida2018, Shchur2020} for different neuronal models. 
But in the Appendix \ref{app} we are able to prove that any initial distribution converges to a unique  
stationary distribution given by (\ref{fs}).

\begin{figure}
\centering
\includegraphics[width=0.9\textwidth]{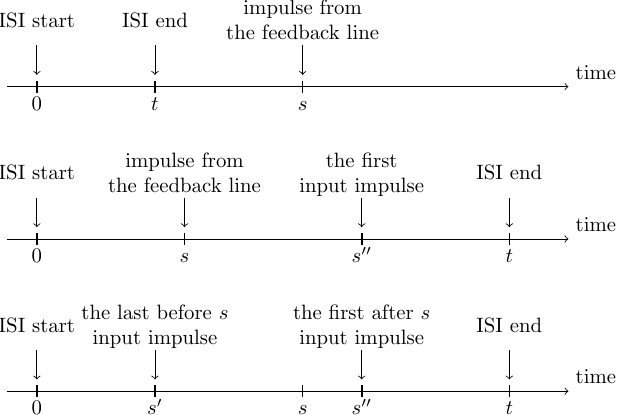}
\caption{\label{s's''figure} To calculate the conditional PDF $p(t|s)$ to get the output ISI of a duration $t$ if at the beginning
of the ISI there was an impulse in the feedback line with 
time-to-live $s$, one should consider three possible mutually exclusive events.  
Top panel: an event ``to obtain the first from the ISI beginning output impulse at $t < s$ ''. Middle panel: an event ``to obtain the first output impulse at $t>s$ if the first input impulse arrives after at $s''>s$ ''.
Bottom panel: an event  ``to obtain the first output impulse at the moment $t>s$ if the first input impulse arrives before $s$ ''. $s'$ and $s''$ are the arrival times  of  two consecutive input impulses, namely, the last one before and the first one after obtaining the impulse from  the feedback line. Points $0$ and $t$ on all three time lines denote
the start and the end of the ISI, respectively, and $s$ denotes the arrival time of the impulse from the feedback line.}
\end{figure}

%In order to obtain $p(t)$, we still need the conditional PDF $p(t|s)$.  
To find $p(t|s)$, let us consider three different cases, see Fig. \ref{s's''figure}. In Fig. \ref{s's''figure}, points $0$ and $t$ on the time lines denote
the start and the end of the ISI, respectively. Note that at the beginning of each ISI, i.e. at the time moment $0$, an input impulse has been received, and it triggers a neuron. Therefore at the moment $0$, the input stochastic process starts anew, because it is renewal. Further, at that moment, 
the neuron is in its resting state, and there is an inhibitory impulse in the feedback line with time-to-live $s \in [0; \Delta ]$.

The first possible case is to get an output ISI $t < s$, see top panel
of Fig. \ref{s's''figure}. The latter means that the impulse from the feedback line does not reach the neuron before the end of the ISI. Thus the probability $p(t)dt$ to obtain an ISI of duration within  $[t;t+dt[$ for a neuron with feedback is the same as for the 
same neuron without feedback, i.e. 
\begin{equation}\nonumber
p(t|s)=p^0(t),\quad t < s.
\end{equation}

The second possible case is to obtain an output ISI of duration $t>s$. 
Then  two mutually exclusive scenarios should be considered. The
first one is when there is no input impulses  during the first $s$ time units after the start of the ISI, or, in other words, before the arrival of the impulse from the feedback line, see middle panel
of Fig. \ref{s's''figure}. An event ``to obtain the first output impulse at the moment $t>s$ if the first input impulse arrives after the one from the feedback line'' constists of the continuum of the alternative events 
indexed with a parameter $s''\in ]s;t[$. $s''$ denotes the arrival time of the first input impulse after the start of  the ISI.
Each alternative event consists of the following two consecutive  and statistically independent events:
\begin{enumerate}
\item The first input impulse  is  obtained  at the time $s'' \in ]s;t[$. The event has the probability $p^{in}(s'')ds''$. 

\item The output impulse is produced at the time $t>s''$.
At the moment $s''$, the neuron appears in the same state as it will be at the beginning of an ISI in case of instantaneous excitatory feedback described in \cite{Vidybida2015a}. After the moment $s$, the delayed feedback inhibition does not have any influence on the neuron. The input stochastic process also starts anew at $s''$. Thus the probability density to get the output ISI of duration $t$ is $p^{o\_if}(t-s'')$,
where $p^{o\_if}(t)$ is the PDF for the same neuron but  with instantaneous
excitatory feedback instead of inhibitory one. We use here notation from \cite{Vidybida2015a}.
\end{enumerate}

Therefore the probabilty of the considered alternative event indexed with the  parameter $s''$ is $p^{in}(s'')p^{o\_if}(t-s'')ds''dt$. Its contribution to $p(t|s)$ is as follows:
\begin{equation}\nonumber
p(t|s)=\int\limits_s^t ds''\:p^{in}(s'')
p^{o\_if}(t-s''),\quad t>s.
\end{equation}

The other scenario in case of $t>s$
is an event ``to obtain the first output impulse at the moment $t>s$ if the first input impulse arrives before the one from the feedback line''. It consists of the continuum of the alternative events 
indexed with two parameters $s'\in ]0;s]$ and $s''\in ]s;t[$, see bottom panel of Fig. \ref{s's''figure}. $s'$ and $s''$ are the time moments of obtaining two consecutive  impulses from the input line during one ISI, namely the last one before and the first one after obtaining the impulse from  the feedback line, respectively. Each alternative event consists of three consecutive  and statistically independent events:
\begin{enumerate}
\item The last impulse from the input line before the time moment $s$ reaches the neuron at the moment $s'<s$, and the neuron has not been triggered during the time interval $]0;s']$. 
Note that during the time interval $]0;s']$ the neuron is not being influenced by the feedback and therefore can be considered as the one without feedback. We denote the probability density of such events (obtaining an input impulse at the moment 
$s'$ before obtaining the impulse from the feedback line and the absence of firings during the time interval $]0;s']$) as
$\tilde{P}^0(s')$.

\item The next impulse after the input impulse at $s'$ is  obtained from the input at the moment $s''>s$. The probability density of such events is $p^{in}(s''-s')$.
It does not depend on the neuronal state or previous input impulses because the input stream is expected to be renewal. 

\item The neuron is trigerred at the time $t>s''$. At the moment $s''$, right after the receiving the input impulse,  the neuron  is in the same state as at the beginning of an ISI
in case of instantaneous excitatory feedback.  Therefore the probability density to get the output ISI of duration $t$ is $p^{o\_if}(t-s'')$.
\end{enumerate}

Thus  the probability of the considered alternative event indexed with the parameters $s'$ and $s''$ is $\tilde{P}^0(s')p^{in}(s''-s')p^{o\_if}(t-s'')ds'\: ds''dt$. 

Finally, after examining three mutually exclusive events, the conditional PDF $p(t|s)$ can be written as follows:
\begin{equation}\label{pts}
p(t|s)=\chi(s-t)p^0(t)+\int\limits_s^t ds''\:
p^{o\_if}(t-s'')\left(p^{in}(s'')+ \int\limits_0^s ds'
{\tilde P}^0(s') p^{in}(s''-s')\right),
\end{equation}
where $\chi(s-t)$ denotes a Heaviside step function.

In \cite{Vidybida2015a}, the relation between the ISIs PDF for a neuron without feedback $p^0(t)$
and the ISIs PDF for the same neuron with instantaneous excitatory feedback $p^{o\_if}(t)$ was derived:
\begin{equation}\label{p0ifgen}\nonumber
p^0(t)=\int\limits_0^t dt'\: p^{in}(t') p^{o\_ if}(t-t').
\end{equation}

After applying the Laplace transform over the last expression,
the Laplace transform of  $p^{o\_if}(t)$
can be obtained:
\begin{equation}\label{laplacep0ifgen}
\mathcal{L}\{p^{o\_ if(t);s}\}=\dfrac{\mathcal{L}\{p^0(t);s\}}{\mathcal{L}\{p^{in}(t);s\}}.
\end{equation}

In order to calculate $p(t)$, the exact expression for the distribution $\tilde{P}^0(s')$ is needed. It can be found as follows. The probability density of not trigerring a neuron without feedback
during the first $t'$ time units of an ISI is   a complementary cumulative distribution function (CCDF) of ISIs for the neuron without feedback $P^0(t')$:
\begin{equation}\label{P0}
P^0(t')=1-\int\limits_0^{t'} dt\: p^0(t).
\end{equation}
The event of not trigerring the neuron
during the first $t'$ time units of an ISI can be represented as a sum of two mutually exclusive events. The first one is an event 
``to obtain no input impulses during the first $t'$ time units of an ISI''. Its probability density is the CCDF for the input stream:
\begin{equation}\label{Pin}
P^{in}(t')=1-\int\limits_0^{t'} dt''\: p^{in}(t'').
\end{equation}

The second event is ``to not trigger a neuron
during the first $t'$ time units of an ISI if at least one input impulse is fed into the neuron during that time''.  It can be represented as a set of alternative events indexed with a parameter $s'\in ]0;t']$. The alternative event consists of
 two consecutive independent events. The first event is that
at the moment $s'$ the neuron obtains an input impulse and has not been triggered during the time interval $]0;s']$. The probability of such an event
is $\tilde{P}^0(s')ds'$. The second event is the absence of input impulses  during the time interval $]s';t']$. Note that at the moment $s'$  the input stochastic process starts anew. Thus the probability of the second event is $P^{in}(t'-s')dt'$.

Therefore  $\tilde{P}^0(s')$
can be derived by solving the following integral equation:
\begin{equation}\label{tildeP}
P^0(t')=P^{in}(t')+\int\limits_0^{t'} ds'\: {\tilde P}^0(s')P^{in}(t'-s').
\end{equation}

Performing the Laplace transform over the last equation and using Eqs. (\ref{P0}) and (\ref{Pin}), one can obtain the Laplace transform of $\tilde{P}^0(s)$:
\begin{equation}\label{laplceP0tilde}
\mathcal{L}\{{\tilde P}^0(t);s\}=\dfrac{\mathcal{L}\{p^{in}(t);s\}-\mathcal{L}\{p^0(t);s\}}{1-\mathcal{L}\{p^{in}(t);s\}}. 
\end{equation}

Substituting Eq. (\ref{pts}) for the conditional probability $p(t|s)$ into Eq. (\ref{pt}) and taking into account that $f(s)$ can be represented as in Eq. (\ref{fs}), the
following expressions for the ISIs PDF $p(t)$ for the neuron with delayed feedback inhibition can be derived:

\begin{multline}\label{ptfin1}
p(t)=
\int\limits_0^t ds\: g(s) \int\limits_s^t ds''\:
p^{o\_if}(t-s'')\Bigg(p^{in}(s'')+\\
+
\int\limits_0^s ds'{\tilde P}^0(s')
 p^{in}(s''-s')\Bigg)+p^0(t)\Bigg(\int\limits_t^\Delta ds\: g(s)
+a\Bigg),\quad t<\Delta;
\end{multline}
\begin{multline}\label{ptfin2}
p(t)=
\int\limits_0^\Delta ds\: g(s)
\int\limits_s^t ds''\: p^{o\_if}(t-s'') \left(p^{in}(s'')+
\int\limits_0^s ds'{\tilde P}^0(s')
 p^{in}(s''-s')\right)+\\
 +a\int\limits_\Delta^t ds''\:p^{o\_if}(t-s'')\left( p^{in}(s'')+\int\limits_0^\Delta ds'{\tilde P}^0(s')
 p^{in}(s''-s')\right),\quad
t>\Delta.
\end{multline}

Notice that, according to Eqs. (\ref{ptfin1}) and (\ref{ptfin2}), the obtained ISIs PDF $p(t)$ has a jump type discontinuity at a point corresponding to the output ISI that is equal to the delay in the feedback line $\Delta$:
\begin{equation}\nonumber
\lim_{t\rightarrow \Delta^-} p(t)-\lim_{t\rightarrow \Delta^+} p(t)=a \: p^0(\Delta).
\end{equation}

To summarize, if the output ISIs PDF for the same neuron without feedback $p^0(t)$ and the input ISIs PDF $p^{in}(t)$ are known, then the output ISIs PDF $p(t)$ for the neuron with delayed feedback inhibition  can be calculated as follows:
\begin{enumerate}
\item to find the stationary PDF of times-to-live $f(s)=g(s)+a\:\delta(\Delta-s)$, first, calculate $g(s)$ using Eq. (\ref{g}), then find $a$ from the normalization
condition on $f(s)$;  
\item reverse the Laplace transform of $p^{o\_ if}(t)$ in Eq. (\ref{laplacep0ifgen});
\item reverse the Laplace transform of $\tilde{P}^0(t)$ in Eq. (\ref{laplceP0tilde});
\item substitute all mentioned above into Eqs. (\ref{ptfin1}) and (\ref{ptfin2}) to calculate $  p(t)$.
\end{enumerate}

The results obtained here for the renewal input 
have been checked analytically that they
are applicable to the most simple case of  the renewal process, namely,  a Poisson point process, studied previously in \cite{Vidybida2018}.

\subsection{Model invariant segment of $p(t)$ for neurons with threshold 2 stimulated with Erlang-2 stream}\label{sect_erlang}
To illustrate the obtained in previous section general formulas, we will apply them to a case when the input stream is renewal, but non-Poissonian. 

Firstly,
consider a neuronal model without feedback that satisfies
conditions specified in Sect. \ref{class} above. 
Input ISIs are assumed to be distributed according to the Erlang-2 distribution:
\begin{equation}\label{erlang}
p^{in}(t)=\lambda^2 t e^{-\lambda t},\quad \lambda>0.
\end{equation}

Hereinafter it is assumed that the neuron has threshold 2. It means that  2 input impulses may trigger
the neuron provided they are close in time. This can be realized, for instance, for  the standard leaky integrate-and-fire model satisfying the  condition
$h < V_0 <2h$, where $V_0$ is the firing threshold, and $h$ is the height of stimulating impulses.

In the considered case of the neuron with threshold 2,
there is a domain of ISIs  $t \in ]0;T_2[$ such that the second after the ISI start input impulse 
received at the moment $t$ will definitely trigger an output spike. For instance, for the binding neuron model $T_2=\tau$ \cite{Vidybida2008}, where $\tau$ is the storage time of the internal memory. For the mentioned above  leaky integrate-and-fire neuron model \cite{Vidybida2017}, 
\begin{equation*}
T_2=\tau \ln \frac{h}{V_0-h},
\end{equation*}
here $\tau$ is the membrane relaxation time.

From the definition of $T_2$, it follows that for the ISIs of a duration less than $T_2$ time units,
the event of the neuron trigerring consists of two consecutive independent events of
obtaining an input impulse. Therefore, in such a case, for the neuron without feedback the ISIs PDF $p^0(t)$ is as follows:
\begin{equation}\label{p0erlang}
p^0(t)=\lambda e^{-\lambda t} \dfrac{(\lambda t)^3}{3!},\quad t<T_2.
\end{equation}

The latter together with Eq. (\ref{erlang})  allows a calculation of the PDF ${\tilde P}^0(s')$ for $s'<T_2$. Since Eq. (\ref{laplceP0tilde}) is valid, the distribution ${\tilde P}^0(s')$ is given by the following:
\begin{equation}\label{P0tildeerlang}
{\tilde P}^0(s')=p^{in}(s'),\quad s'<T_2.
\end{equation}

Further, for the ISIs
shorter than $T_2$, the event of triggering the neuron with instantaneous
excitatory feedback is the event of
obtaining an input impulse (one impulse is already fed into the neuron at the beginning of an ISI due to the instantaneous feedback). Thus the output ISI
PDF for the neuron with instanteneous excitatory feedback
$p^{o\_if}(t)$ is the same as the input ISIs PDF $ p^{in}(t)$ (\ref{erlang}):
\begin{equation}\label{poiferlang}
p^{o\_if}(t)=p^{in}(t), \quad t<T_2.
\end{equation}

Secondly, consider the case when the delayed inhibitory feedback line is added to the neuron discussed right above. Assume that 
the delay in the feedback line $\Delta$ fulfills the following condition:
\begin{equation}\label{condition}
\Delta<T_2.
\end{equation}

Note that, to find the PDF of times-to-live $f(s)$, the explicit expression for $p^0(t)$ is needed only for the arguments within the interval $[0;\Delta ]$. Thus, if the condition (\ref{condition})  holds, the PDF of times-to-live $f(s)=g(s)+a\:\delta (\Delta-s)$ can be found in the considered case of the neuron with threshold 2 stimulated with the
Erlang-2 stream. After the substitution of the expression   (\ref{p0erlang}) for $p^0(t)$ into Eqs.
(\ref{g}), (\ref{g0}) of the Appendix \ref{app}, where the expression for $g(s)$ is given,  one can obtain the following:
\begin{equation}\label{gerlang}
g(s)=
 \dfrac{a\lambda  e^{-\lambda (\Delta-s)}}{2 }(\sinh (\lambda (\Delta- s))-\sin (\lambda (\Delta- s))), \quad s \in [0;\Delta],
\end{equation}
where
\begin{equation}\label{aerlang}
a=\dfrac{8}{2e^{-\lambda\Delta}
\left(\cos \left(\lambda\Delta\right)+\sin \left(\lambda\Delta\right)\right)+
2 \lambda\Delta+e^{-2 \lambda\Delta}+5}.
\end{equation}

Finally,  the expressions for $ p^{in}(t)$ (\ref{erlang}),
$p^0(t)$ (\ref{p0erlang}), ${\tilde P}^0(t)$ (\ref{P0tildeerlang}), $p^{o\_if}(t)$ (\ref{poiferlang}), 
 and $g(s)$ (\ref{gerlang}) can be substituted into Eq. (\ref{ptfin1}) to find the PDF for the neuron
with delayed inhibitory feedback $p(t)$ for the ISIs $t<\Delta$:
\begin{equation}\begin{split}\label{pt_erlang2_1}
p(t)&=
\dfrac{a\lambda  e^{-\lambda (2 \Delta+t)}}{2880 }
\Bigg(-45 (2+ \lambda t)+45 e^{2  \lambda t} \Big(2+ \lambda  t\big(-3+\lambda t  (1+ \lambda t )\big)\Big)+\\
&+
\lambda ^2 t^2\Big(45+ \lambda t \big(75+2 e^{2  \lambda \Delta  } (150-30 \lambda  t +60  \lambda \Delta + \lambda ^3 t^3)\big)\Big)+\\
&+60
e^{ \lambda \Delta  } \bigg(e^{\lambda t }  \lambda ^3 t^3 \cos \big( \lambda(t-\Delta ) \big)+3 \lambda  t (-4+\lambda ^2 t^2 ) \cos ( \lambda \Delta
)-\\
& -3 e^{\lambda t } \Big(-4+ \lambda t \big(4+\lambda t   (-2+\lambda t )\big)\Big) \sin \big(\lambda (t-\Delta ) \big)+\\
&+\Big(12+\lambda ^2 t^2 (-6+\lambda t 
)\Big) \sin (\lambda \Delta   )\bigg)\Bigg),
\end{split}\end{equation}
and into Eq. (\ref{ptfin2}) to find $p(t)$ for the ISIs $t \in ]\Delta;T_2[$:
\begin{equation}\begin{split}\label{pt_erlang2_2}
p(t)&=\frac{a\lambda e^{-2 \lambda (t+\Delta )  }}{2880}   
\Bigg(15 e^{\lambda t } \Big(-6+ \lambda t \big(-3+ \lambda  t (3+5  \lambda t)\big)\Big)+\\
&+e^{\lambda (t+2 \Delta
) } \bigg(90+5 \lambda ^3 t^3 \big(45+2   \lambda \Delta (3+  \lambda \Delta ) (9+2 \lambda  \Delta )\big)+\\
&+60 \lambda \Delta \Big(-15+\lambda \Delta 
  \big(15+\lambda \Delta   (10+  \lambda \Delta)\big)\Big)-\\
&-15\lambda^2   t^2 \Big(3+2 \lambda  \Delta   \big(-15+\lambda \Delta    ( \lambda \Delta   (10+\lambda \Delta
  )+15)\big)\Big)+\\
& +3  \lambda t \Big(255+2   \lambda \Delta \big(\lambda  \Delta (5+  \lambda \Delta) (-45+  \lambda \Delta (15+2   \lambda \Delta))-135 \big)\Big)\bigg)+\\
 &+60
e^{\lambda(t+\Delta )  } \Big(3  \lambda t (-4+ \lambda ^2 t^2) \cos (  \lambda \Delta)+\big(12+\lambda ^2 t^2  (-6+ \lambda t)\big)
\sin (  \lambda \Delta)\Big)\Bigg),
\end{split}\end{equation}
where the coefficient $a$ is given by Eq. (\ref{aerlang}).

\begin{figure}
\centering{
\includegraphics[width=\textwidth]{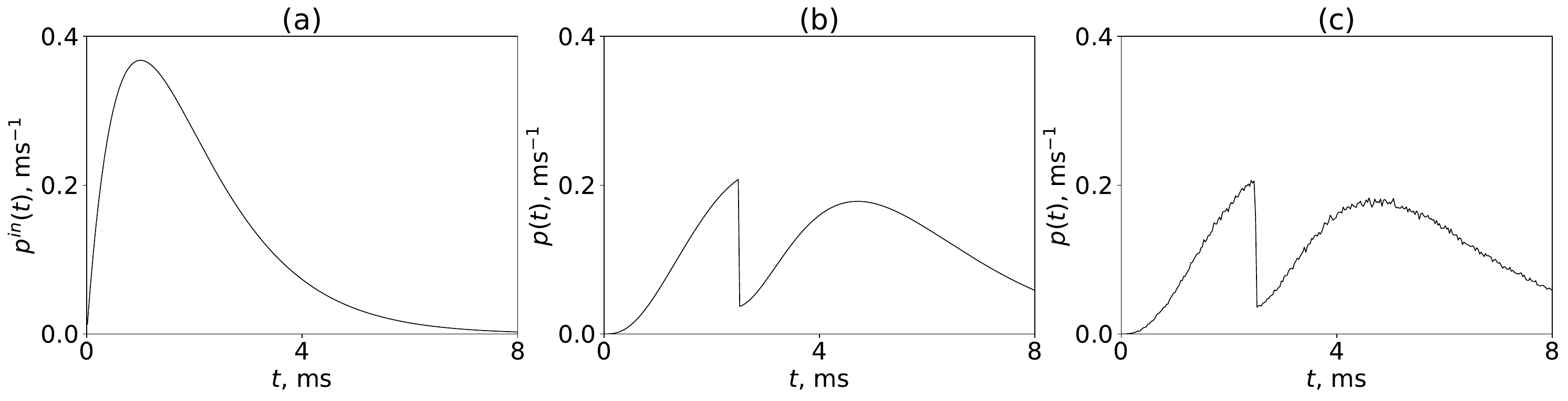}
}
\caption{\label{examples}  Example of the ISIs PDF for the input
Erlang-2 stream (a), Eq. (\ref{erlang}) is used, and for the output stream of the binding neuron with threshold 2 with delayed inhibitory  feedback, stimulated with the  Erlang-2 stream (b), Eqs. (\ref{pt_erlang2_1}) and (\ref{pt_erlang2_2}) are used.
  Neuronal model parameters used for both panels:
$\tau=8$ ms, $\lambda=1$ ms$^{-1}$, and $\Delta=2.5$ ms in (b).
(c): results of the Monte Carlo simulation for the same neuron as in (b) (1\,000\,000 output ISIs were obtained).
% {\color{red} The total probability under the curves} is 0.996981 (a) and 0.905041 (b), (c).
 As it can be seen from the plot (b), the delayed feedback inhibition causes a jump type discontinuity of the ISIs PDF at a point corresponding to the output ISI that equals to the delay in the feedback line $\Delta$. The analytical results for the neuron with feedback inhibition (b) are in concordance with the results of the numerical simulation (c).}
\end{figure}

Notice that expressions for the ISIs PDF $p(t)$ (\ref{pt_erlang2_1}) and
(\ref{pt_erlang2_2}) have been obtained without specifying the exact
neuronal model.
For example, we might consider the leaky integrate-and-fire neuron, the perfect integrator, or the binding neuron model. It means that at some initial interval  of ISI durations, the PDF $p(t)$
is the same for the whole considered in this section class of  neuronal models with
delayed inhibitory feedback. The further course of $p(t)$ will
depend on the distribution for the neuron without feedback $p^0(t)$, which is different for different neuronal models.

The expressions (\ref{pt_erlang2_1}) and
(\ref{pt_erlang2_2}) for $p(t)$ have been checked numerically by means of the
Monte Carlo simulation for the binding neuron model with threshold 2
stimulated with the Erlang-2 stream (\ref{erlang}), and with delayed inhibitory feedback
specified in Sect. \ref{class}. The model parameters have been set to fulfill the conditions
(\ref{condition}) and (\ref{f(s)condition}). The results of the simulation are displayed in Fig. \ref{examples}.

\section{Discussion}
The consideration of a renewal process as a neuronal input in comparison with a Poisson point process, as it is done in \cite{Vidybida2018}, poses a separate challenge and requires its own approach. This is due to the fact that for a Poisson process time intervals between events are distributed according to the exponential distribution. The exponential distribution is the only continuous probability distribution that is memoryless \cite[Sect. I.3]{Feller1971}. The memorylessness greatly simplifies the study of a neuronal activity.

In the present paper, the obtained analytical results are valid if the condition (\ref{f(s)condition}) is fulfilled,
which ensures the existence of the unique stationary distribution of times-to-live $f(s)$. Actually, the condition (\ref{f(s)condition}) is  equivalent to the requirement that the input ISIs smaller than
 $\Delta$ have low probability. This condition is in line with biological details of the synaptic transmission.
Indeed, when a synapse mediates impulses at high frequency,
i.e. short input ISIs are highly probable, the time of the synaptic 
transmission can be prolonged up to the hundreds of ms due to the asynchronous synaptic release.
It was shown that the latter may happen to autapses of PV neurons \cite{Manseau2010}. However,
in the current paper, we assume that 
the impact of an inhibitory impulse from the feedback line on a neuron is momentary (see Sect. \ref{class} above).  

\bigskip

{\small
\textit{Acknowledgments}. This work was supported by the Programs of Basic Research of the Department of Physics and Astronomy of the National Academy of Sciences of Ukraine ``Mathematical models of nonequilibrium processes in open systems'', № 0120U100857, and ``Noise-induced dynamics and correlations in nonequilibrium systems'',
№ 0120U101347.

\appendix
\section{The convergence of  $f(s)$ to a stationary distribution} \label{app}
Suppose that there is an ensemble of identical neurons.
Each of them has an impulse in the feedback line with some value of time-to-live $s$. Also, suppose that, at the beginning, these times-to-live $s$ have a distribution $f_0(s)$. It is clear that the distribution of $s$ can be changed only after triggering. Denote 
the distribution obtained after the $n$-th trigerring of every neuron in the ensemble as $f_n(s)$. Here we prove that $f_n(s)$ converges\footnote{The regular part of $f_n(s)$ converges in $C([0;\Delta])$
and the singular part (the $\delta$-function mass) converges in $R^1$. }
 to a distribution $f(s)$ which remains the same after further triggerings (is stationary).

In the work \cite{Vidybida2015}, to find the distribution of times-to-live $f(s)$, the transition function $\mathbf{P}(s | s' )$ was introduced. It gives the probability density to find an impulse in the feedback line at the beginning
of an ISI  with time-to-live $s$ provided that, at the beginning of the previous
ISI, there was an impulse with time-to-live $s'$. Thus, if the distribution of times-to-live at the beginning of an ISI was $f_n(s)$, then at the beginning of the next ISI
it is given by
\begin{equation}\begin{split}\label{fequat_appendix}
f_{n+1}(s)&=\int\limits_0^\Delta ds'\:\mathbf{P}(s | s' ) f_n(s')=\\
&=\int\limits_s^\Delta ds'\: p^0(s'-s)f_n (s')+
\delta(s-\Delta )\int\limits_0^\Delta ds'\: P^0(s')f_n(s'),
\end{split}\end{equation}
where the explicit form of $\mathbf{P}(s | s' )$ was taken into account. 

Since $f_n(s)$ is a PDF, it is normalized:
 \begin{equation}\nonumber
\int\limits_0^ \Delta ds\: f_n(s) = 1.
\end{equation}

With the help of Eq. (\ref{fequat_appendix}), it can be checked that a norm of the
distribution of times-to-live at the beginning of each ISI does not change:
\begin{equation}\label{normsave}
\int\limits_0^ \Delta ds\: f_{n+1}(s) = \int\limits_0^ \Delta ds\: f_n(s).
\end{equation}

According to Eq. (\ref{fequat_appendix}), $f_n(s)$ can be represented as follows:
\begin{equation}\nonumber
f_n(s)=g_n(s) +a_n \delta(s-\Delta ),
\end{equation}
where $g_n(s)\in C([0;\Delta])$, and $a_n \in [0;1]$. Then Eq. (\ref{fequat_appendix}) can
be rewritten as the following system of equations on $g_n(s)$ and $a_n$:
\begin{equation}\label{fscases}
\begin{cases}
a_{n+1}=\int\limits_0^\Delta ds'\: P^0(s')g_n(s')+a_n P^0(\Delta);\\
g_{n+1}(s)=\int\limits_s^\Delta ds'\: p^0(s'-s)g_n(s')+ a_n p^0(\Delta-s).
\end{cases}
\end{equation}

However, since Eq. (\ref{normsave}) is valid, $a_n$ can be determined through $g_n(s)$:
\begin{equation}\nonumber
a_n = 1- \int\limits_0^ \Delta ds\: g_n(s),
\end{equation}
which, after substitution into the second equation of the system (\ref{fscases}), gives
the self-contained expression for the sequence $\{g_{n}(s)\}$:
\begin{equation}\label{gn}
g_{n+1}(s)=\int\limits_s^\Delta ds'\: p^0(s'-s)g_n(s')+ p^0(\Delta-s)(1- \int\limits_0^ \Delta ds'\: g_n(s')).
\end{equation}

Let us assume that $p^0(s)\in \mathcal{F} = C([0;\Delta]),\;s\in [0;\Delta]$. Then Eq. (\ref{gn}) can be rewritten in terms of an operator $M:\mathcal{F}\rightarrow \mathcal{F}$:
\begin{equation}\nonumber
(M g)(s)=\int\limits_s^\Delta ds'\: p^0(s'-s)g(s')+ p^0(\Delta-s)(1- \int\limits_0^ \Delta ds'\: g(s')).
\end{equation}

If we introduce  an operator $\tilde{M}:\mathcal{F}\rightarrow \mathcal{F}$  such as $M=Inv\; \tilde{M}\; Inv$, where  $(Inv \; g)(s)=g(\Delta-s)$, then the previous equation can be rewritten as follows:
\begin{equation}\label{tildeMg}\nonumber
( \tilde{M} g)(s)=\int\limits_0^s ds'\: p^0(s-s') g(s')+ p^0(s)(1-\int\limits_0^ \Delta ds'\: g(s')).
\end{equation}

If the condition
\begin{equation}\label{f(s)condition}
\int\limits_0^\Delta p^0(s)ds+\Delta\sup_{s \in [0; \Delta]}p^0(s)<1
\end{equation}
is fulfilled, then the operator $\tilde{M}:\mathcal{F}\rightarrow \mathcal{F}$ is a contraction
in $C([0;\Delta])$. Consequently, the sequence $\{g_n(s)\}$ defined by $g_{n+1}(s)
=(\tilde{M} g_n)(s)$ 
converges to a unique fixed point of the operator $\tilde{M}$. This proves the existence and uniqueness of the PDF
$f(s)$ in the stationary regime. The stationary $g(s)$ is a solution of the following equation:
 \begin{equation}\label{geq}
g(s)=\int\limits_0^s ds'\: p^0(s-s') g(s')+ p^0(s)(1-\int\limits_0^ \Delta ds'\: g(s')).
\end{equation}

The solution of Eq. (\ref{geq}) 
is given by \cite[p. 631]{polyanin2008handbook}:
\begin{equation}\label{g}
g(s)=\dfrac{g^0(s)}{1+\int\limits_0^ \Delta g^0(s')ds'},
\end{equation}
where 
\begin{equation}\label{g0}
g^0(s)=\sum_{k=0}^\infty (V^k p^0)(s),
\end{equation}
and an operator $V$ is defined as follows:
\begin{equation}\nonumber
(V \phi )(s) = \int\limits_0^s ds'\: p^0(s-s') \phi (s'),\quad
 \phi \in \mathcal{F}.
\end{equation} 
It can be easily shown that $||V||<1$ in $C([0;\Delta])$, which ensures convergence in (\ref{g0}).

\end{document}